# Matter Waves and Orbital Quantum Numbers

*Roger Ellman*

Abstract

The atom's orbital electron structure in terms of quantum numbers (principal, azimuthal, magnetic and spin) results in space for a maximum of: *2* electrons in the *n=1* orbit, *8* electrons in the *n=2* orbit, *18* electrons in the *n=3* orbit, and so on. Those dispositions are correct, but that is not because of quantum numbers nor angular momentum nor a "Pauli exclusion principle".

Matter waves were discovered in the early 20th century from their wavelength, which was predicted by DeBroglie to be, Planck's constant divided by the particle's momentum, $\lambda_{mw} = h/m \cdot v$. But, the failure to obtain a reasonable theory for the matter wave frequency resulted in loss of interest. That problem is resolved in "A Reconsideration of Matter Waves" [2] in which a reinterpretation of Einstein's derivation of relativistic kinetic energy [which produced his famous $E = m \cdot c^2$] leads to a valid matter wave frequency and a new understanding of particle kinetics and the atom's stable orbits.

It is analytically shown that the orbital electron arrangement is enforced by the necessity of accommodating the space that each orbiting electron's matter wave occupies.

Roger Ellman, The-Origin Foundation, Inc.
    320 Gemma Circle, Santa Rosa, CA 95404, USA
    RogerEllman@The-Origin.org
    http://www.The-Origin.org

# Matter Waves and Orbital Quantum Numbers

*Roger Ellman*

In general the 20th Century concept of the overall arrangement of atom's orbital electrons developed as follows.

- The electron orbits are located in *shells,* a shell conceptually being a spherical surface with the atomic nucleus located at the center of the sphere. The locations available for electron orbits are in a series of concentric shells corresponding to the orbit number, $n$, referred to as the principal quantum number, $n$ taking the integer values $1, 2, 3, \ldots$.

– Each shell may have a set of *sub-shells*. An additional quantum number, $\ell$, an integer and referred to as the azimuthal quantum number, is defined. It may have each of the integer values in the range $0$ to $n-1$, each value corresponding to a separate sub-shell.

- The electron's *orbital angular momentum is spatially quantized.* This refers to permitted relative tilts among the electron orbits of a shell. The spatial quantization is that the projection of the angular momentum vector on an axis of measurement can only be certain integral multiples of $[h/2 \cdot \pi]$. The orbital angular momentum in a shell is $\ell \cdot [h/2 \cdot \pi]$.

A third quantum number, $m_\ell$, an integer and called the orbital magnetic quantum number, is defined so that $m_\ell$ may take the integer values from $+\ell$ to $-\ell$, a total of $[2 \cdot \ell + 1]$ values. The allowed projections of the angular momentum on a selected axis of measurement are each of the allowed values of $m_\ell \cdot [h/2 \cdot \pi]$, where the various projections differ because of different tilting of the various orbits.

- A characteristic *spin* is attributed to the orbital electron. Its angular momentum may only have the value $\frac{1}{2} \cdot [h/2 \cdot \pi]$. Depending upon whether the spin angular momentum vector is in the same or the opposite direction as the orbital angular momentum vector a fourth "spin" quantum number, $m_s$, has the value $\pm$ that angular momentum.

- The Pauli Exclusion Principle operates; that is: no two electrons in the same atom may have identical values for all four of the above quantum numbers, $n$, $\ell$, $m_\ell$, and $m_s$.

Included in this conception of the orbital electrons are that the orbits may be elliptical as well as circular and that the orbital electron is conceived of as not so much an object in a specific location as an effect "smeared out" over a substantial portion of the orbit. Generally, the above shells concept of the orbital structure of multi-electron atoms is validated in its agreement with the spectral data, the chemical behavior characteristics and the Periodic Table of the Elements.

Application of this set of rules results in the set of available locations for electrons in an atom as listed in Table 1, on the following two pages.

*Table 1*
*"Quantum Number" Description of Orbital Electrons Arrangements*

| Element Name | Z | Electron Number | n | ℓ | m_ℓ | m_s |
|---|---|---|---|---|---|---|
| Hydrogen | 1 | 1 | 1 | 0 | 0 | −½ |
| Helium | 2 | 1 | 1 | 0 | 0 | −½ |
|  |  | 2 | 1 | 0 | 0 | +½ |
| Lithium | 3 | 1 | 1 | 0 | 0 | −½ |
|  |  | 2 | 1 | 0 | 0 | +½ |
|  |  | 3 | 2 | 0 | 0 | −½ |
| Beryllium | 4 | 1 | 1 | 0 | 0 | −½ |
|  |  | 2 | 1 | 0 | 0 | +½ |
|  |  | 3 | 2 | 0 | 0 | −½ |
|  |  | 4 | 2 | 0 | 0 | +½ |
| Boron | 5 | 1 | 1 | 0 | 0 | −½ |
|  |  | 2 | 1 | 0 | 0 | +½ |
|  |  | 3 | 2 | 0 | 0 | −½ |
|  |  | 4 | 2 | 0 | 0 | +½ |
|  |  | 5 | 2 | 1 | −1 | −½ |
| Carbon | 6 | 1 | 1 | 0 | 0 | −½ |
|  |  | 2 | 1 | 0 | 0 | +½ |
|  |  | 3 | 2 | 0 | 0 | −½ |
|  |  | 4 | 2 | 0 | 0 | +½ |
|  |  | 5 | 2 | 1 | −1 | −½ |
|  |  | 6 | 2 | 1 | −1 | +½ |
| Nitrogen | 7 | 1 | 1 | 0 | 0 | −½ |
|  |  | 2 | 1 | 0 | 0 | +½ |
|  |  | 3 | 2 | 0 | 0 | −½ |
|  |  | 4 | 2 | 0 | 0 | +½ |
|  |  | 5 | 2 | 1 | −1 | −½ |
|  |  | 6 | 2 | 1 | −1 | +½ |
|  |  | 7 | 2 | 1 | 0 | −½ |
| Oxygen | 8 | 1 | 1 | 0 | 0 | −½ |
|  |  | 2 | 1 | 0 | 0 | +½ |
|  |  | 3 | 2 | 0 | 0 | −½ |
|  |  | 4 | 2 | 0 | 0 | +½ |
|  |  | 5 | 2 | 1 | −1 | −½ |
|  |  | 6 | 2 | 1 | −1 | +½ |
|  |  | 7 | 2 | 1 | 0 | −½ |
|  |  | 8 | 2 | 1 | 0 | +½ |

*Table 1 [continued]*
*"Quantum Number" Description of Orbital Electrons Arrangements*

| Element Name | Z | Electron Number | n | $\ell$ | $m_\ell$ | $m_s$ |
|---|---|---|---|---|---|---|
| Fluorine | 9 | | | | | |
| | | 1 | 1 | 0 | 0 | $-\frac{1}{2}$ |
| | | 2 | 1 | 0 | 0 | $+\frac{1}{2}$ |
| | | 3 | 2 | 0 | 0 | $-\frac{1}{2}$ |
| | | 4 | 2 | 0 | 0 | $+\frac{1}{2}$ |
| | | 5 | 2 | 1 | -1 | $-\frac{1}{2}$ |
| | | 6 | 2 | 1 | -1 | $+\frac{1}{2}$ |
| | | 7 | 2 | 1 | 0 | $-\frac{1}{2}$ |
| | | 8 | 2 | 1 | 0 | $+\frac{1}{2}$ |
| | | 9 | 2 | 1 | 1 | $-\frac{1}{2}$ |
| Neon | 10 | | | | | |
| | | 1 | 1 | 0 | 0 | $-\frac{1}{2}$ |
| | | 2 | 1 | 0 | 0 | $+\frac{1}{2}$ |
| | | 3 | 2 | 0 | 0 | $-\frac{1}{2}$ |
| | | 4 | 2 | 0 | 0 | $+\frac{1}{2}$ |
| | | 5 | 2 | 1 | -1 | $-\frac{1}{2}$ |
| | | 6 | 2 | 1 | -1 | $+\frac{1}{2}$ |
| | | 7 | 2 | 1 | 0 | $-\frac{1}{2}$ |
| | | 8 | 2 | 1 | 0 | $+\frac{1}{2}$ |
| | | 9 | 2 | 1 | 1 | $-\frac{1}{2}$ |
| | | 10 | 2 | 1 | 1 | $+\frac{1}{2}$ |
| Sodium | 11 | | | | | |
| | | 1 | 1 | 0 | 0 | $-\frac{1}{2}$ |
| | | 2 | 1 | 0 | 0 | $+\frac{1}{2}$ |
| | | 3 | 2 | 0 | 0 | $-\frac{1}{2}$ |
| | | 4 | 2 | 0 | 0 | $+\frac{1}{2}$ |
| | | 5 | 2 | 1 | -1 | $-\frac{1}{2}$ |
| | | 6 | 2 | 1 | -1 | $+\frac{1}{2}$ |
| | | 7 | 2 | 1 | 0 | $-\frac{1}{2}$ |
| | | 8 | 2 | 1 | 0 | $+\frac{1}{2}$ |
| | | 9 | 2 | 1 | 1 | $-\frac{1}{2}$ |
| | | 10 | 2 | 1 | 1 | $+\frac{1}{2}$ |
| | | 11 | 3 | 0 | 0 | $-\frac{1}{2}$ |
| Magnesium | 12 | | | | | |
| | | 1 | 1 | 0 | 0 | $-\frac{1}{2}$ |
| | | 2 | 1 | 0 | 0 | $+\frac{1}{2}$ |
| | | 3 | 2 | 0 | 0 | $-\frac{1}{2}$ |
| | | 4 | 2 | 0 | 0 | $+\frac{1}{2}$ |
| | | 5 | 2 | 1 | -1 | $-\frac{1}{2}$ |
| | | 6 | 2 | 1 | -1 | $+\frac{1}{2}$ |
| | | 7 | 2 | 1 | 0 | $-\frac{1}{2}$ |
| | | 8 | 2 | 1 | 0 | $+\frac{1}{2}$ |
| | | 9 | 2 | 1 | 1 | $-\frac{1}{2}$ |
| | | 10 | 2 | 1 | 1 | $+\frac{1}{2}$ |
| | | 11 | 3 | 0 | 0 | $-\frac{1}{2}$ |
| | | 12 | 3 | 0 | 0 | $+\frac{1}{2}$ |

[and so forth.]

The problem with this conception of the orbital electrons is that it does not address the question of how the electrons are behaving and why they are doing so.

In the paper "A Reconsideration of Matter Waves"[2] it as shown that the Bohr hypothesis is actually that the length of each stable orbital path must be an integral number of orbital electron matter wavelengths.  A reason is presented for that matter wavelength restriction.  The question therefore arises:  what reasons or causes impel the orbital electrons into their structure in multi-electron atoms and just what is that structure ?

The orbital electron extends a distance of $½·\lambda_{mw}$ forward and rearward of its instantaneous location (its location is the locus of its Coulomb action).  In effect the orbital electron occupies that much space.  The space that the matter wave occupies is like a long narrow tube.  The "tube" is straight and tangential to the electron's location on its orbital path, that is, the orientation of the matter wave "points" in the direction of the orbital electron's instantaneous velocity.

There are three constraints that govern the behavior of the orbital electrons:

(1) The orbital path length must be an integral number of matter wavelengths, as already developed.

(2) The electrons being all of the same charge magnitude and polarity, tend to repel each other to a spacing equally apart subject to the common central attraction of the oppositely charged nucleus.

(3) The electron spacing along the orbital path must be such that the $½·\lambda_{mw}$ extension of the electron in space forward and rearward of its current position does not interfere with the space correspondingly occupied by any of the other electrons.

Of course, in addition there are the obvious constraints that the number of electrons in orbit must be the same as the number of equivalent positive charges in the nucleus because the atom is overall electrically neutral and that the electron orbits and the electron positions in the orbits must be such that they do not collide nor otherwise interfere with each other.

The orbital electron arrangements of the above Table 1 result in there being space for a maximum of:  *2*  electrons in the  *n=1*  orbit,  *8*  electrons in the  *n=2*  orbit,  *18*  electrons in the  *n=3*  orbit, and so on.  Those dispositions are correct; but that is not because of "quantum numbers" nor angular momentum nor a "Pauli exclusion principle".  That orbital electron arrangement is enforced by the requirement of accommodating the space that each orbiting electron's matter wave occupies, as follows.

Applying the constraints to the innermost  *n=1*  orbit where the orbital path length is  $\lambda_{mw}$  there is only space for two electrons in the orbital plane [see Figure 1, on the following page and equation *(1)*, below].  In the figure the second electron is depicted located as close to the first electron as possible without their matter wave extensions in space interfering with each other.  Introduction of a third electron into that orbit in that plane would involve spacing that would disrupt the particles and the orbit.  Since there can only be two electrons in the orbit and they repel each other they will space  *180°*  apart.

```
(1)  For the n = 1 orbit or "shell" the orbital path
     length, the circular path circumference, is one
     wavelength, 2π·R = λ_mw.   Then from Figure 2, below:

                ½·λ_mw      ½·2π·R
     Tan(Φ)  = ────────  = ────────  =  π
                  R            R

     Φ  =  72.34°

     Electron Space  =  360°/2·Φ  =  2.49  ⇨  2 electrons
```

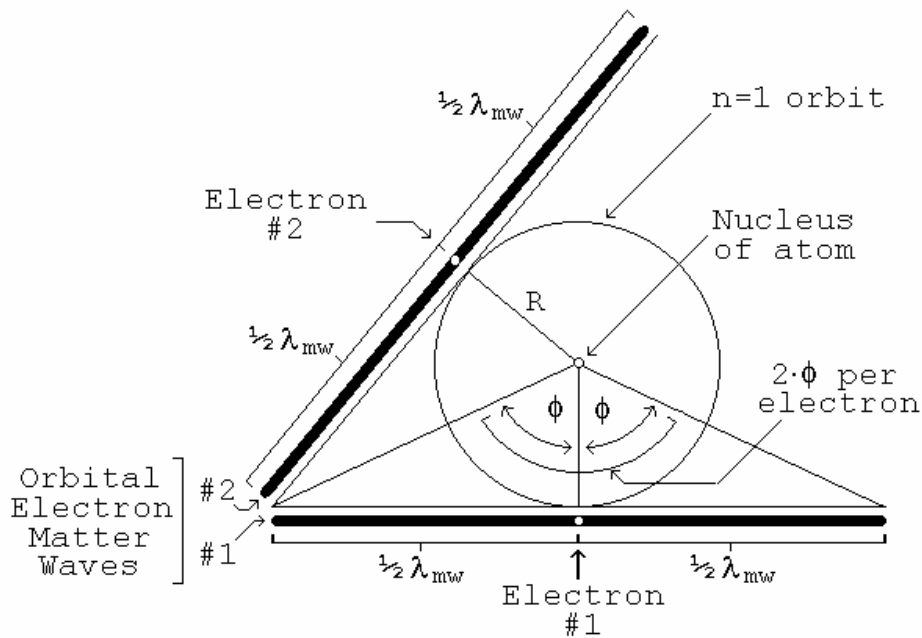

*Figure 2*
*Electrons in n=1 Shell*

    Now considering adding a third electron in a second $n=1$ orbit with its orbital plane tilted relative to the orbit of the (above) first two electrons, the situation is somewhat like that of a sword dance where a number of dancers whirl and turn, each flashing a pair of swords, one in each hand, while avoiding any casualties among the dancers. The dancers' spacing, paths and timing must be such that while their swords slash at each others' paths they do so when the dancer in that path (with his extended swords) is at another location on the path.

    If a second two-electron orbit is introduced in a plane tilted relative to the above first $n=1$ orbital plane the third electron will interfere with the first two regardless of the tilt of its orbital plane relative to the other. This is readily seen by imagining in Figure 2, above, that the paper is folded along the line from the nucleus to where the two matter waves are shown just meeting. The fold tilts one electron's orbit relative to the other's but does not change the interference of the two. Thus, in terms of the angles in Figure 2, a second orbital plane tilted at an angle of $\Phi = 72.34°$ or more would seem to fit.

    However, the electron in that second orbital plane, starting at $\Phi = 72.34°$ above one of the points of intersection with the first plane could travel only the distance $[180° - 2 \cdot \Phi] = 35.32°$ before being within $\Phi = 72.34°$ of the other side of the orbit, the other point of intersection of the planes. During that $35.32°$ the pair of electrons in the original plane have not had the necessary travel, $\Phi = 72.34°$, to clear their matter wave extensions in space from the common points of intersection of the two orbital planes.

    Therefore, the $n=1$ shell can only contain one orbital plane with only one orbit having two equally spaced electrons. Any additional content would involve the matter waves of the electrons interfering with each other. The "dancers slashing swords" would at least clash if not injure the dancers.

    For the $n=2$ orbit the sword dance becomes more complex. Clearly, from the above, the first two $n=2$ electrons can readily share an orbit as in the $n=1$ case. In fact, calculation analogous to equation *(1)* but for the $n=2$ case shows that three electrons could fit in one $n=2$ orbital plane. That calculation appears at equation *(2)*, below.

*(2)* For the n = 2 orbit or "shell" the orbital path
length, the circular path circumference, is two
matter wavelengths, $2\pi \cdot R = 2 \cdot \lambda_{mw}$ or $\lambda_{mw} = \pi \cdot R$.

$$\tan(\Phi) = \frac{\tfrac{1}{2} \cdot \lambda_{mw}}{R} = \frac{\tfrac{1}{2} \cdot \pi \cdot R}{R} = \pi/2$$

$\Phi = 57.52°$

Electron Space = $360°/2 \cdot \Phi$ = 3.13 ⇨ 3 electrons

However, the fit is close and more overall equidistant spacing of the electrons is achieved with the third electron occupying a new orbital plane tilted to the first as develops below.

How many such tilted planes can be accommodated at the *n=2* level in total? The shell can accommodate three such planes at *θ = 60°* relative tilts. This limit is set by $\Phi_{n=2}$ = *57.52°*. Four planes tilted at *θ = 45°* would be too close. The three planes have a common axis of intersection on which are the two points that all three of the orbits have in common (Figure 3, below).

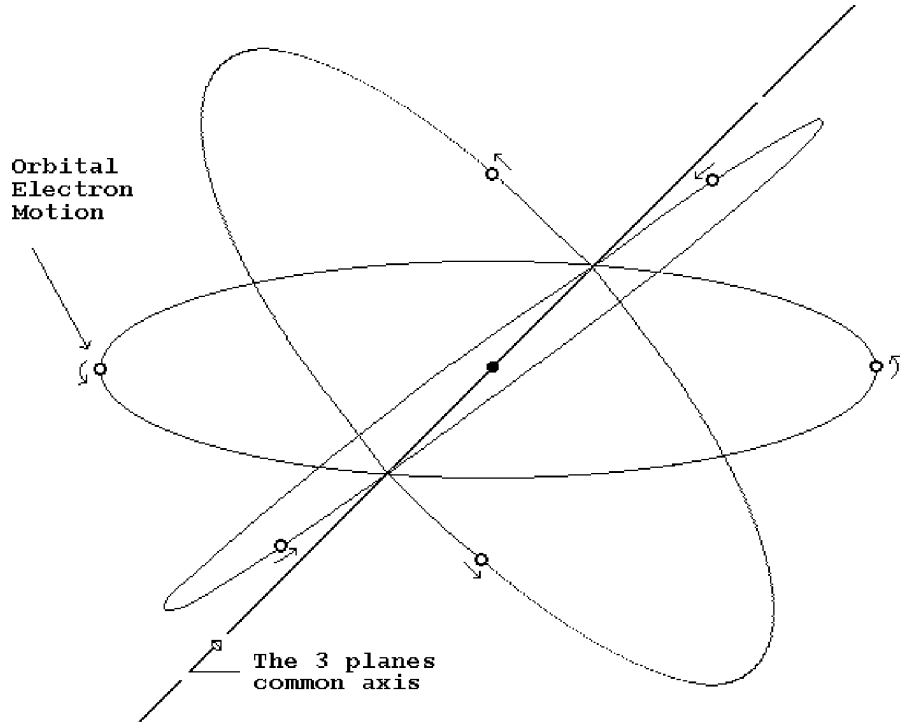

*Figure 3*
*Three Orbital Planes and Relative Tilts, n=2 Shell*

The six electrons (two per each of three orbital planes tilted at *60°* relative to each other) pass through those two common points at *φ = 360°/6 = 60°* intervals (equidistant spacing). With *Φ = 57.52°* there is just enough travel between successive electrons for each electron to clear the area before the next one starts arriving.

Now the reason for only two electrons in each of the orbital planes here, even though three could fit in any one such plane, becomes clear. With three electrons per plane the electrons (all evenly spaced) would pass through the two common points of the three orbital planes every *φ = 360°/9 = 45°*. That is closer than the minimum $\Phi_{n=2}$ = *57.52°* spacing required in this *n=2* shell of orbits.

Can any more electrons fit in this shell ? Yes, two more in another orbital plane perpendicular to the common axis of the other three orbital planes. This new orbit intersects each of the other three successively at $\Theta = 60°$ intervals. The two electrons in each such intersected plane are spaced $180°$ apart. An electron passing such an intersection with one of the first three planes $60°$ after one of that plane's two electron's has passed and taking $60°$ to clear the intersection would have cleared the requisite $60°$ ahead of the other electron of that plane. Two such electrons $180°$ apart can be accommodated. Overall, therefore the number of orbital electrons that can fit in the $n=2$ shell is eight: two in each of the three planes depicted in Figure 3, above, plus two more in the plane perpendicular to the axis of those first three planes.

For $n=3$ the situation becomes considerably more complex. Now the separation angle is $\Phi_{n=3} = 46.32°$. The reasoning as for $n=2$, above, indicates that the shell can still accommodate only three orbital planes intersecting on a common axis, each plane having two electrons in orbit $180°$ apart with the one more plane perpendicular to the common axis of the other three planes. In other words, for $n=3$ the shell appears able to only accommodate the same orbital structure as does the $n=2$ shell. This is in fact the case.

More precisely, the $n=3$ shell so functions until full in that form. Additional electrons for higher $Z$ atoms then start filling the $n=4$ shell. Then, the electric field of those outer $n=4$ electrons becomes sufficient to modify the orbital structure situation and possibilities of the inner $n=3$ shell. The $n=3$ shell then can accommodate the expected five orbital planes on a common axis, each with two electrons, in addition to the already filled $n=2$ type structure. For higher $n$ the same kind of effect of outer on inner shell modifies the structure, the $n=5$ shell filling partly before the $n=4$ shell is completely filled and that partial outer shell's field then modifying the inner shell's structure.

It is the complex fitting of the space occupied by the orbital electron matter waves into the available integer-matter-wavelength orbital shells that determines the orbital electrons' arrangement structure. That structure is summarized in Table 4, below. The table, arranged so as to directly correspond to the quantum number system of 20th Century physics presented on the first page of this paper, shows what those quantum numbers actually represent and why they are able to produce correct results.

But it should be observed that the way in which the requirements imposed by the orbital electron matter waves force the structure of the atom's electron orbits has nothing to do with angular momentum and has nothing to do with quantization. Any effects observed in the atomic orbital structure that have the appearance of quantization are merely the result of fundamental and simple mechanical spacing requirements operating.

In other words, the entire structural effect is the result of the matter waves of the orbital electrons and the restrictions that their space requirements impose on the system. While the appearance of quantization of angular momentum is there in some forms and with various modifications and adjustments (such as projections on an axis), that is only because of the relationship between angular momentum and matter wave length; that is, that a statement of quantized angular momentum is actually a statement of integer values of matter wavelength.

The statement that the orbital electron's angular momentum is quantized, as in the following traditional equation

$$(3) \quad m \cdot v \cdot R = n \cdot \frac{h}{2\pi} \quad\quad\quad [n = 1, 2, \ldots]$$

is merely a mis-arrangement of

(4)  $2\pi \cdot R = n \cdot \dfrac{h}{m \cdot v} = n \cdot \lambda_{mw}$     [n = 1, 2, ...]

a statement that the orbital path length, $2\pi \cdot R$, must be an integral number of matter wavelengths, $n \cdot \lambda_{mw}$, long. The latter statement has a clear, simple, operational reason for its necessity. The former statement is arbitrary and is justified only because it produces the correct result, even if without an underlying rational reason.

Of course, one no longer needs a "Pauli Exclusion Principle". Rather, it is clear that if two orbital electrons had the same four quantum numbers the two electrons would be co-located, an obvious impossibility for orbital electrons.

*Table 4*
*The Orbital Structure Significance of "Quantum Numbers"*

| Quantum Number | Orbital Structure |
|---|---|
| n | The "index number" of the "shell". The shell's orbital path length is "n" matter wavelengths long.<br><br>n = 1, 2, 3, … |
| $\ell$ | The "index number" of the particular "set" of orbital planes in the "shell".<br><br>$\ell$ = 0, 1, … n-1<br><br>A "set" consists of orbital planes of orbits of the same length, tilted at equal angles relative to each other, and sharing the same common axis about which tilted.<br><br>The number of "sets" in a particular "shell" equals [$\ell$ + 1]. |
| $m_\ell$ | The "index number" of any particular orbital plane in any particular "set" of orbital planes.<br><br>$m_\ell$ = +[$\ell$], +[$\ell$ − 1], … 0, −1, … −[$\ell$]<br><br>The total number of such orbital planes in the "set" is<br><br>[$2 \cdot \ell$ + 1], always odd. |
| $m_s$ | Each individual orbital plane can accommodate 2 electrons equally spaced. [While for n > 1 more than two electrons could be accommodated in any one plane of a set, for the planes of the set taken together only two electrons per plane can be accommodated.]<br><br>$m_s$ = −½ and +½ [for the 1st and 2nd electrons of the plane]. |

*References*

[1] This paper is based on development in R. Ellman, *The Origin and Its Meaning*, The-Origin Foundation, Inc., http://www.The-Origin.org, 1997, in which the development is more extensive and the collateral issues are developed.

[2] R. Ellman, "A Reconsideration of Matter Waves," http://www.arXiv.org, physics/9808043.